\documentclass[letterpaper]{article} 
\usepackage{aaai2027}  
\usepackage[hyphens]{url}  
\usepackage{graphicx} 
\usepackage{natbib}  
\usepackage{caption} 
\usepackage{amsmath}
\usepackage{amssymb}
\usepackage{algorithm}
\usepackage{algorithmic}
\usepackage{multirow}

\title{SEER: A Scalable Event-Driven Runtime for Multi-Agent Reinforcement Learning Environment Execution}
\author{
    Jiming Su,
    Feng Zhu,
    Yuhao Xiao,
    Miao Zhang
}
\affiliations{
    NUDT\\
    jimings@nudt.edu.cn,
}

\begin{document}

\maketitle
\begin{abstract}
Multi-agent reinforcement learning (MARL) is increasingly used in systems where many entities interact in a shared environment. In these workloads, policy inference can be batched on GPUs, but the CPU-side environment path still updates entity states, resolves local interactions, and builds observations inside each $\mathrm{env.step}$. This path can dominate latency in interaction-intensive workloads after policy batching. Existing vectorized environment interfaces and general event-driven runtimes do not directly address the fine-grained cross-entity work inside one bounded MARL step while preserving the step-level semantics expected by learning systems. We present SEER (Scalable Event-Driven Environment Runtime), a runtime for MARL environment execution under an explicit event-locality contract. SEER models each environment step as a bounded stream of timestamped events. If direct state updates are entity-local and cross-entity effects are represented as ordered future events, SEER can run independent events in parallel while preserving serial-equivalent boundary outputs. To make this contract practical, SEER uses a CAS-based dual-layer event queue: local event queues preserve per-entity causal order, while a parallel executive list exposes ready work to global worker threads. SEER further uses STPA, a preflight-guided thread-pool selection mechanism that combines short runtime traces with static workload and hardware features. We evaluate SEER on two parameterized MARL workloads: compute-intensive UAV swarm control (Aircraft) and communication-intensive wireless management (PCS). On Aircraft, SEER reduces inference latency by 16.2\% over the baseline. On PCS, SEER reduces environment-execution latency by 46.9\%, 19.5\%, and 64.9\% over ROSS-T, NBSE, and ROOT-Sim, respectively. These results suggest that fine-grained event scheduling and workload-aware thread selection can reduce MARL environment-execution overhead in the evaluated interaction-intensive workloads.
\end{abstract}

Multi-agent reinforcement learning (MARL) increasingly supports large shared environments, including robot swarms, wireless networks, transportation systems, and digital twins~\cite{vinyals2019grandmaster,jaderberg2019human,luong2019applications,orr2023multiagent}. After GPU policy batching, CPU-side state updates, interaction processing, and observation construction can become the dominant remaining cost~\cite{weng2022envpool}. We study how to parallelize this intra-environment execution within one shared MARL step.

Vectorized environments parallelize independent instances but do not expose irregular interaction work inside one shared $\mathrm{env.step}(A_t)$~\cite{terry2021pettingzoo,weng2022envpool}. Such work includes entity updates, collision or contention handling, message processing, and observation construction. The key problem is therefore to expose intra-step parallelism while preserving the learning-loop outputs for environments that satisfy an explicit event-locality contract.

\begin{figure}[!t]
    \centering
    \includegraphics[width=1\linewidth]{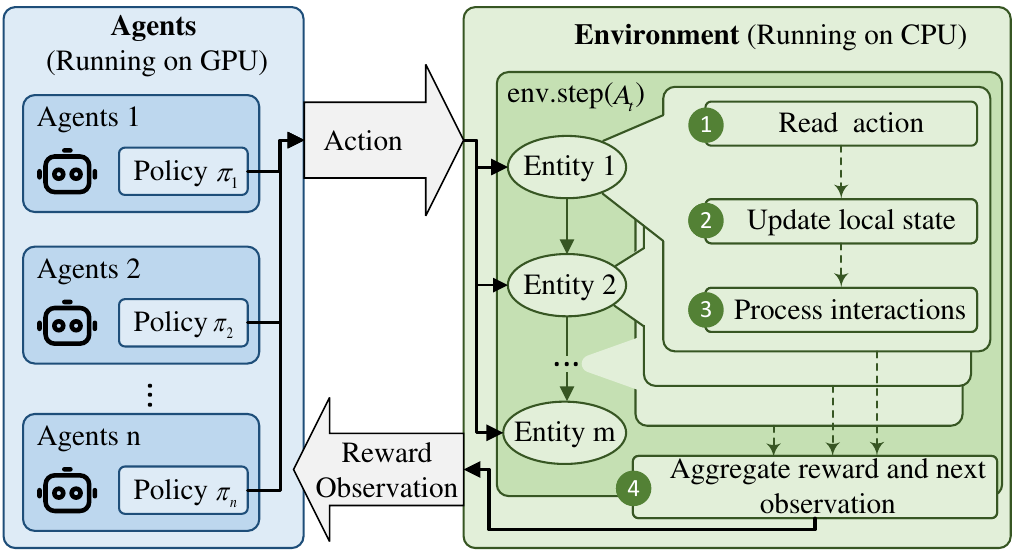}
    \caption{MARL environment-execution workflow.}
    \label{fig:1}
\end{figure}

Fig.~\ref{fig:1} shows the execution pattern. After agents produce actions, the CPU updates entity states, processes interactions, and constructs rewards and observations. Conventional traversal preserves determinism but serializes many entity-local operations that could execute concurrently.

Generic parallel discrete-event simulation does not directly satisfy the learning-facing boundary of a MARL step: the next rewards, termination flags, and observations must preserve the original environment semantics. SEER therefore adopts an event-locality contract in which direct writes are entity-local, cross-entity effects are represented as ordered events, and outputs are emitted only after the in-step event window completes.

SEER separates semantic order, work sharing, and resource selection. Entity-local queues preserve causal order, a global ready-event list exposes executable work, and preflight-guided thread selection chooses a suitable degree of parallelism. Unlike environment-level vectorization or open-ended PDES, SEER targets bounded execution inside one shared MARL step.

This paper makes the following contributions:

\begin{itemize}
    \item We identify intra-environment execution as a major post-batching bottleneck and define an event-locality contract for entity-causal parallel execution.
    \item We design a CAS-based dual-layer queue and dynamic scheduler that separate local causal order from global ready-work sharing.
    \item We use preflight traces and static workload features to predict a suitable thread-pool size for short MARL executions.
    \item We evaluate throughput, latency, load balance, thread prediction, and component contributions on Aircraft and PCS against ROSS-T, NBSE, and ROOT-Sim.
\end{itemize}

\section{Problem Model and Motivation}

This section defines the event model used by SEER, measures where time is spent in representative MARL inference steps, and turns these observations into design requirements.

\subsection{Event-Based Model of MARL Environment Execution}

We model the CPU path as a bounded event-processing problem. The environment contains $N$ entities $\mathcal{L}=\{L_1,\ldots,L_N\}$ with states $s_i\in\Sigma_i$, while $M$ agents produce joint action $A_t$. The environment maps $A_t$ to entity inputs and transforms $(S_t,A_t)$ into $(S_{t+1},R_t,D_t,O_{t+1})$ through events:

\begin{equation}
e = \langle \tau, \iota, \eta, \pi, \mathrm{seq} \rangle,\quad
\tau \in \mathcal{T},\; \iota \in \{1,\ldots,N\},\;
\eta \in \mathcal{H},\; \mathrm{seq} \in \mathbb{N}.
\end{equation}

Here, $\tau$ is the timestamp, $\iota$ is the target entity, $\eta$ is the event type, $\pi$ is the payload, and $\mathrm{seq}$ is a deterministic global sequence number. SEER orders events by $\kappa(e)=(\tau(e),\mathrm{seq}(e))$ and stores the pending events of each entity in $Q_i$. The local transition is:

\begin{equation}
\delta_i : \Sigma_i \times \mathcal{E}_i \rightarrow \Sigma_i \times 2^{\mathcal{E}}.
\end{equation}

where $\mathcal{E}_i$ contains events targeting $L_i$. Function $\delta_i$ updates only $L_i$ and may generate new events. Cross-entity effects must be encoded as future ordered events or explicit conflict-resolution events. Generated events satisfy $\kappa(e')>\kappa(e)$; environments with direct shared writes outside this contract are not covered by SEER's equivalence claim.

The $t$-th environment step is a bounded execution window. SEER encodes each input $u_{i,t}$ as $e_{i,t}^{act}=\langle \tau_t,i,\mathrm{ACT},u_{i,t},\mathrm{seq} \rangle$, executes ready events in $[\tau_t,\tau_{t+1})$, and emits outputs at $\tau_{t+1}$:

\begin{equation}
\mathrm{Step}_t(S_t,A_t)
= \mathrm{Out}_{\tau_{t+1}}\Big(
\mathrm{Exec}_{[\tau_t,\tau_{t+1})}\big(S_t,\; \bigcup_i Q_i \cup \{e_{i,t}^{act}\}_{i=1}^{N}\big)
\Big).
\end{equation}

Here, $\mathrm{Exec}$ consumes in-step events in nondecreasing $\kappa$ order, and $\mathrm{Out}$ executes the boundary output events. Correct parallel execution follows the Local Causality Constraint (LCC)~\cite{fujimoto1990parallel,jefferson1985virtual}:

\begin{equation}
\forall i,\; L_i \text{ consumes events in } Q_i \text{ in nondecreasing order of } \kappa.
\end{equation}

LCC orders events only within the same entity. Under the event-locality contract, concurrently executed events modify disjoint entity states, while every cross-entity effect is materialized as an ordered event before it can update the target entity. Therefore, reordering independent events does not change the state observed at the step boundary. SEER claims serial-equivalent boundary outputs only for workloads that satisfy this contract and complete all in-step events before emitting outputs.

\subsection{Motivation}

\begin{figure}[!t]
    \centering
    \includegraphics[width=1\linewidth]{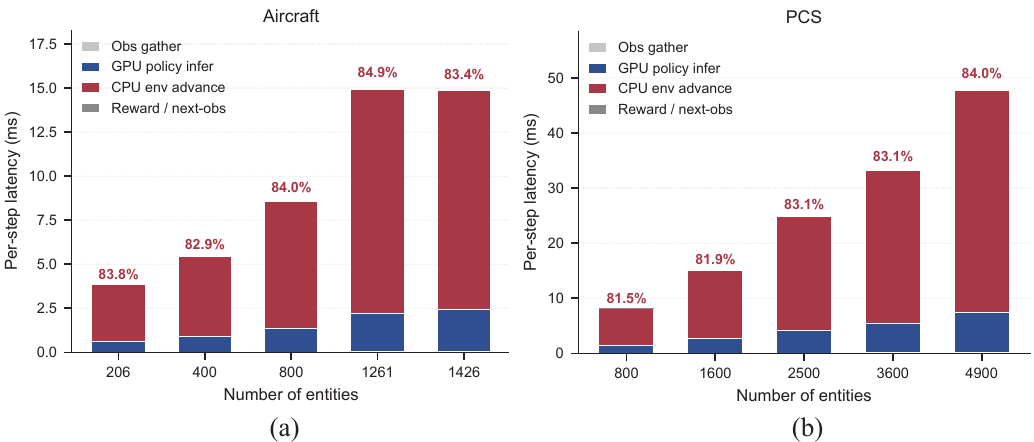}
    \caption{Single-step inference-time distribution in representative MARL environment scenarios. The policy network is MAPPO~\cite{yu2022surprising}, and one agent corresponds to one environment entity.}
    \label{fig:2}
\end{figure}

\textbf{Observation 1.} CPU-side environment execution dominates the measured representative MARL inference steps~\cite{weng2022envpool}. In two MAgent2 scenarios~\cite{zheng2018magent,terry2021pettingzoo}, GPU policy inference in an 800-UAV cooperative-control task takes 1.32 ms, while CPU environment advancement takes 7.26 ms and accounts for 84.0\% of step latency (Fig.~\ref{fig:2}(a)). In a denser wireless-communication scenario, the CPU-side fraction rises from 81.5\% to 86.0\% as the number of entities increases from 800 to 4900. Batched policy inference remains comparatively stable. These measurements show that environment updates and cross-entity interactions are the main remaining bottleneck after policy batching.

\begin{figure}[!t]
    \centering
    \includegraphics[width=1\linewidth]{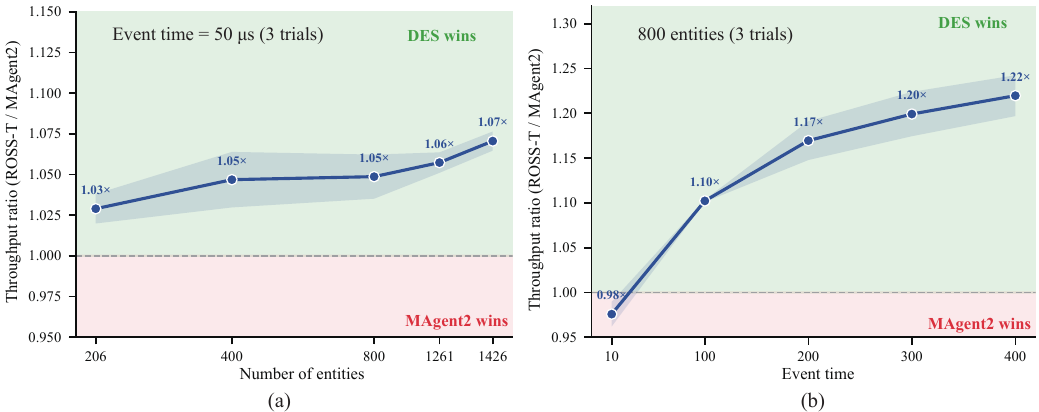}
   \caption{End-to-end MARL inference-throughput ratio between ROSS-T and MAgent2 in the motivation-study workloads. The ratio is computed as ROSS-T throughput divided by MAgent2 throughput under the same motivation-study configuration. Event time denotes the average execution time of events within one environment step. These workloads differ from those used in the main evaluation.}
   \label{fig:3}
\end{figure}

\textbf{Observation 2.} Event-based execution exposes useful intra-step parallelism~\cite{fujimoto1990parallel,weng2022envpool}. In this motivation study, we replace the MAgent2 backend with ROSS-T while keeping the motivation-study scenarios and resource configurations unchanged. Across 10 parameterized configurations, ROSS-T improves end-to-end MARL inference throughput by 1.03--1.22$\times$ over MAgent2 (Fig.~\ref{fig:3}). These experiments are designed only to verify the existence of intra-step parallelism and use workload configurations different from those in the main experimental evaluation. Accordingly, the ratios in Fig.~\ref{fig:3} should not be directly compared with the absolute environment-execution throughput or single-thread-normalized speedups reported in the main evaluation. The event scheduler introduces about 2\% overhead, but its net benefit generally increases with entity count and average event time. This pattern indicates that the environment path contains exploitable parallel work, while also showing that scheduling overhead must be controlled.

Together, the two observations motivate a runtime that is more fine-grained than serial environment stepping, but more workload-aware than directly applying a conventional event-driven runtime. MARL steps are short, hot entities move over time, and cross-entity events appear frequently. A useful runtime must therefore expose parallelism without making queues, locks, or thread oversubscription the new bottleneck.

\subsection{Challenges and Design Goals for Parallel Event Processing}
\textbf{Balancing work while preserving entity order.} Fixed entity-to-thread partitioning improves locality, but it performs poorly when computation-intensive entities shift over time~\cite{fujimoto1990parallel}. Work stealing reduces idle time, but it can add state-migration overhead~\cite{blumofe1999scheduling}. SEER therefore needs global event scheduling under entity-local order.

\textbf{Avoiding queue contention during dynamic scheduling.} A single global queue enables flexible work assignment, but frequent cross-entity events can cause concurrent enqueue/dequeue conflicts and atomic-write contention~\cite{michael1996simple}. Such a queue can also mix same-entity events in an unsafe order. SEER must separate local causal ordering from global work sharing.

\textbf{Predicting the useful thread scale.} Too few threads leave parallelism unused. Too many threads increase context switching, cache invalidation, memory-bandwidth pressure, and synchronization overhead~\cite{hill2008amdahl,pusukuri2011thread,popov2019efficient}. Because the best thread count depends on entity scale, event granularity, interaction structure, and hardware hierarchy, SEER needs low-overhead thread prediction before formal execution begins~\cite{pusukuri2011thread,popov2019efficient,alcaraz2022predicting}.

\section{Design}

\textbf{Key insight.} SEER parallelizes entity-local event execution rather than the entire shared state. Entity queues preserve semantic order, a global ready list exposes parallel work, and preflight traces determine the degree of parallelism.

\subsection{CAS-Based Dual-Layer Event Queue}

\begin{figure}[!ht]
    \centering
    \includegraphics[width=1\linewidth]{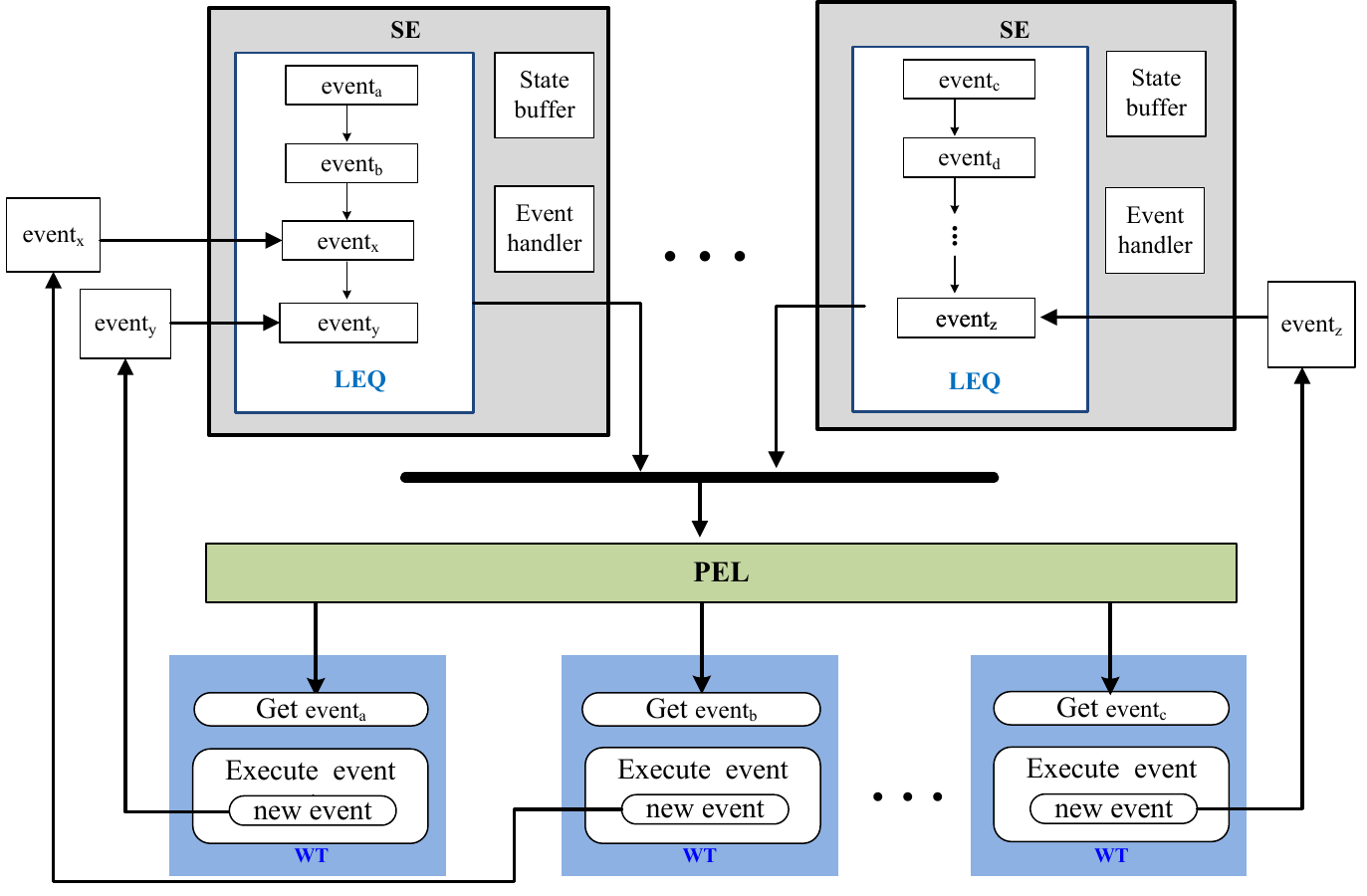}
    \caption{SEER CAS-based dual-layer event queue for MARL environment execution.}
    \label{fig:ldeq}
\end{figure}

SEER uses a distributed CAS-based dual-layer event queue to support fine-grained event scheduling (Fig.~\ref{fig:ldeq}). Each schedulable entity (SE) encapsulates an environment object, owns its state buffer, and consumes its local event sequence in timestamp order. Direct cross-entity state writes are not allowed under the event-locality contract. Cross-entity effects must be expressed as events, which keeps the state boundary explicit and makes parallel execution easier to reason about.

\textbf{Distributed Local Event Queue (LEQ).} Each SE owns an LEQ for events addressed to that entity. The LEQ preserves the entity-local order required by LCC and reduces pressure on a single global structure. Multiple workers may concurrently generate events for the same target entity, while ready-event publication may overlap with insertion; therefore, LEQ operations use atomic synchronization rather than assuming single-threaded access.

\textbf{Parallel Executive Event List (PEL).} Ready events are published from LEQs to the PEL. The PEL contains executable events rather than all pending events. Worker threads therefore compete only for work that can currently run, while local queues continue to enforce per-entity order.

A worker retrieves a PEL event, executes it against the target SE's state buffer, and inserts generated events into the corresponding target LEQs. SE state is not migrated between workers; only event execution is reassigned. This enables global load sharing while preserving entity-local causality, although it may trade some cache locality for better load balance.

\subsection{Dynamic Parallel Scheduling Algorithm}

Algorithm~\ref{Algorithm1} summarizes dynamic scheduling. Each SE maintains ordered events in its LEQ, and only executable events enter the PEL for worker assignment. The runtime drains the PEL by executing events and inserting generated events into target LEQs. When the PEL is empty and all dispatched events at the current timestamp have completed, the runtime advances $t_{cur}$ to the minimum timestamp remaining in the LEQs and republishes the next ready events. The loop preserves LCC because an entity publishes only its next ready timestamp. At the same time, all published events are visible to the worker pool, which lets idle threads process ready work from other entities.

\begin{algorithm}[!h]
    \caption{Dynamic Parallel Event Scheduling.}
    \label{Algorithm1}
    \renewcommand{\algorithmicrequire}{\textbf{Description:}}
    \begin{algorithmic}[1]
        \REQUIRE $N$, number of $SE$s; ${entity}_i$,$0<i\le N$, the $i$-th $SE$; ${event}_{ij}$, the $j$-th environment event of $SE$ ${entity}_i$; $t_{ij}$, timestamp of environment event ${event}_{ij}$; $t_{cur}$, current environment clock; $t_{end}$, environment end clock; ${LEQ}_i$, $LEQ$ for SE ${entity}_i$; $TP$, the thread pool; ${thread}_i$, the $i$-th thread. 
        
\WHILE{$PEL$ is not empty or active workers exist}
    \IF{$PEL$ is not empty and $TP$ is not empty}
        \STATE dequeue an available ${thread}_i$ from $TP$
        \STATE dequeue a ready ${event}_{ab}$ from $PEL$
        \STATE execute ${event}_{ab}$ on ${thread}_i$ and obtain generated event set $\mathcal{G}$
        \FOR{each ${event}_{xy}\in\mathcal{G}$}
            \STATE enqueue ${event}_{xy}$ into ${LEQ}_x$
        \ENDFOR
        \STATE release ${thread}_i$ and enqueue it into $TP$
    \ELSE
        \STATE wait until an active worker completes
    \ENDIF
\ENDWHILE
    \end{algorithmic}
\end{algorithm}

\subsection{Preflight-Guided Thread-Pool Selection}

The best thread-pool size depends on both static workload attributes and dynamic runtime behavior. Static attributes, such as SE count, event granularity, and hardware resources, describe the available parallel work and execution capacity. Dynamic signals, such as contention, critical-section access, context switching, and cache behavior, describe the current cost of using more threads. STPA is a one-shot configuration mechanism for short MARL environment steps. It collects short preflight traces, learns a temporal representation of runtime behavior, fuses this representation with static parameters, and predicts the thread-pool size used for formal execution.

\begin{figure*}[h]
    \centering
    \includegraphics[width=1\linewidth]{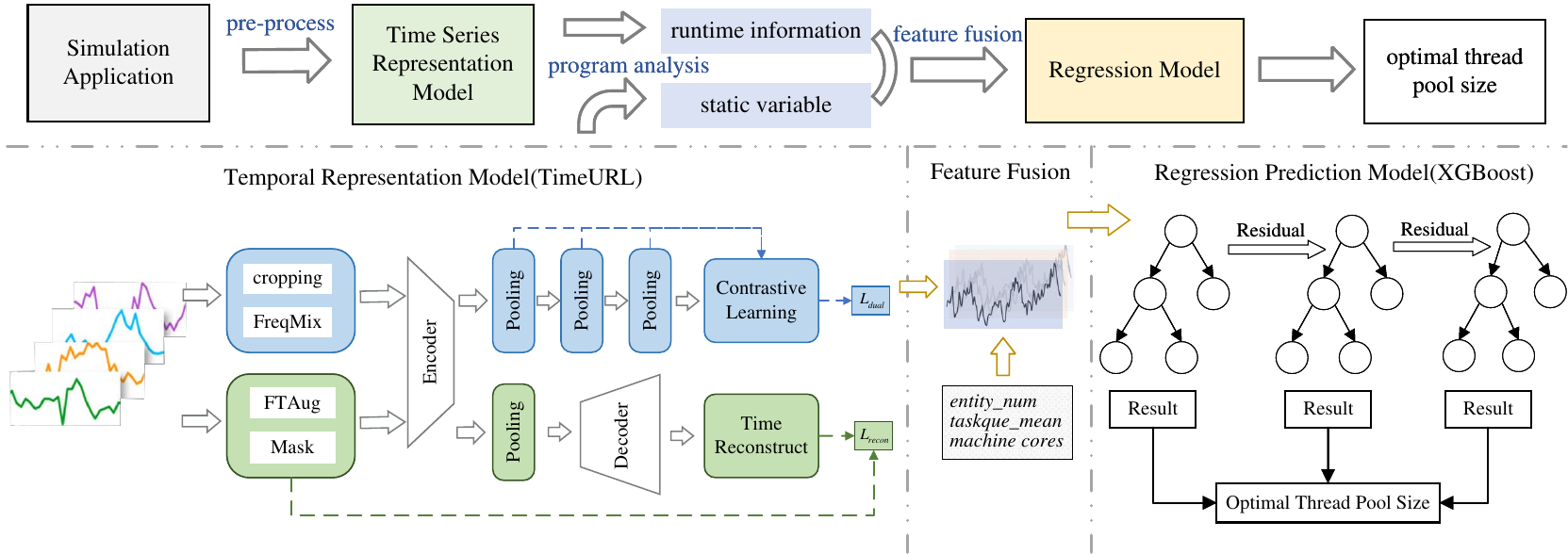}
    \caption{STPA thread-pool prediction for event-driven MARL environment execution.}
    \label{fig:stpa}
\end{figure*}

SEER monitors environment scale, context switching, cache hit rate, and memory consumption. Because per-thread metrics change dimension with the thread-pool size, STPA aggregates runtime metrics by mean, variance, and range, $X(t)=[\mu,\delta^2,R]\in\mathbb{R}^3$. This yields a feature tensor $F\in\mathbb{R}^{N\times T\times 3}$ for $N$ metrics over $T$ preflight steps. Monitoring is non-intrusive: \texttt{pidstat} reads thread-level OS counters, and \texttt{perf} collects PMU/hardware-counter signals. The temporal representation module uses TimesURL~\cite{r48}, and the regression module uses XGBoost~\cite{r49}.


The temporal encoder is trained with contrastive representation learning and masked reconstruction:
\begin{equation}
\mathcal{L}=\frac{1}{|\mathcal{B}|T}\sum_{i,t}(\ell_{temp}^{(i,t)}+\ell_{inst}^{(i,t)})+
\alpha\mathcal{L}_{\text{recon}}.
\end{equation}
The learned representation is passed to XGBoost to predict the normalized thread count. This division keeps the mechanism modular: the temporal encoder summarizes short runtime traces, while the regressor combines that representation with static workload and hardware features. The preflight sampling frequency controls a direct trade-off between feature fidelity and monitoring overhead.

\section{Implementation}

\begin{figure}[!t]
    \centering
    \includegraphics[width=1\linewidth]{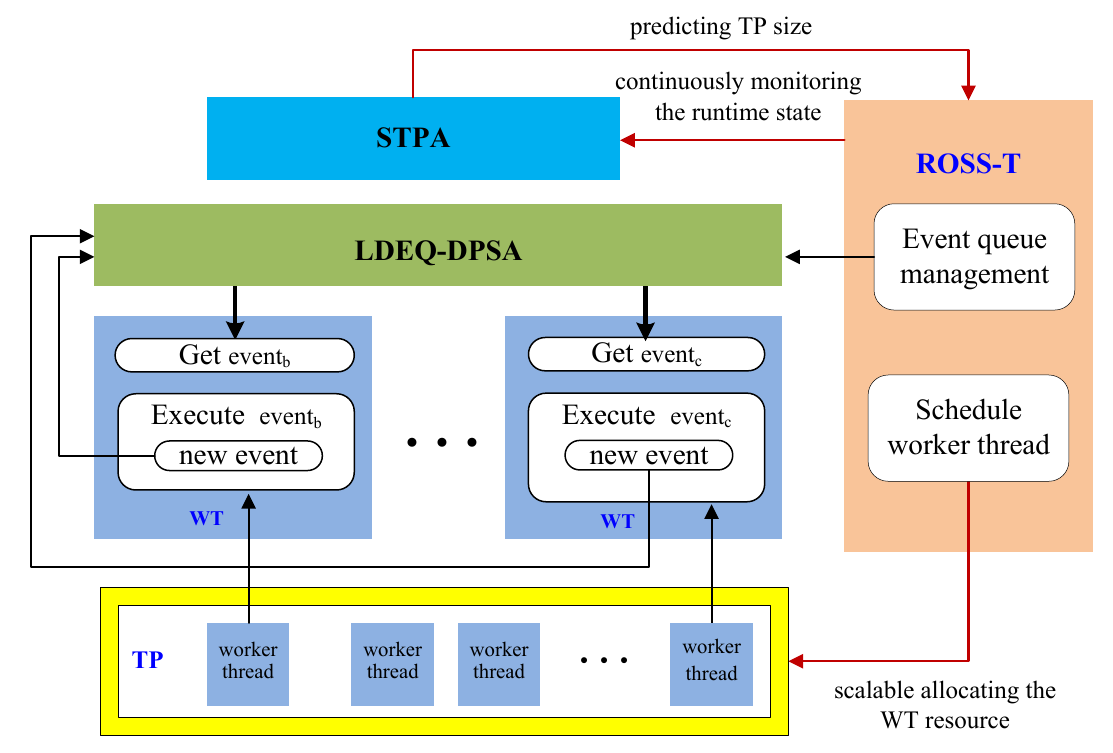}
    \caption{Architecture of the SEER event-driven environment-execution runtime.}
    \label{fig:4}
\end{figure}

We implemented SEER on top of ROSS-T. The implementation contains three modules: the CAS-based dual-layer event queue (LDEQ), the dynamic parallel scheduling algorithm (DPSA), and the preflight-guided thread-pool selection mechanism (STPA), as shown in Fig.~\ref{fig:4}. The modules have separate roles. LDEQ stores ordered local events and publishes ready work. DPSA assigns ready events to worker threads. STPA selects the thread-pool size before formal execution.

\textbf{ROSS-T base runtime.} The baseline is ROSS-T~\cite{jagtap2012optimization}. It binds threads to CPU cores, lets each thread process a private SE task queue, and follows conservative time management. SEER keeps this event-driven execution base but replaces static work binding with the LDEQ-DPSA scheduling path.

\textbf{LDEQ-DPSA.} Events are consumed by multiple worker threads. Instead of lock-based queues, LDEQ-DPSA uses atomic operations, including compare-and-swap, to support concurrent enqueue and dequeue operations under high load~\cite{michael1996simple}. The $Get$ interface retrieves ready events from the PEL, the $Execute$ interface runs event handlers, and generated events are returned to the target LEQs for later publication. Idle workers block on the runtime's worker-queue synchronization primitive and are awakened when new ready events are published, avoiding continuous busy waiting.

\textbf{STPA.} STPA predicts the required thread-pool size from preflight event-arrival rates, system load, and resource utilization. It prevents two common failure modes: insufficient parallelism when too few workers are used, and excessive overhead when too many workers compete for queues and shared resources. STPA assumes that the short preflight phase is representative of the following formal execution. It does not continuously resize the pool within an episode; substantial workload phase changes may therefore reduce prediction quality.

At runtime, SEER forms two coupled loops. The queue-management loop moves ready events from LEQs to the PEL and recirculates generated events. The monitoring-prediction-adjustment loop observes short preflight traces and configures the worker pool. This modular design keeps semantic ordering, work dispatch, and resource selection separate, which makes the runtime easier to extend and evaluate.

\section{Experimental Evaluation}
\label{sec:evaluation}
We evaluate whether SEER improves MARL environment execution under the event-locality contract and without changing the policy network. The evaluation covers two hardware platforms, compute-intensive and communication-intensive workloads, end-to-end inference throughput, environment-execution latency, load balance, thread-selection quality, and ablations.
\subsection{Experimental Design}

\textbf{Hardware platforms.}
We conduct the evaluation on two Linux-based multicore platforms with different processor scales and memory-system characteristics. The two platforms allow us to evaluate SEER under different levels of available hardware parallelism and thread oversubscription. Unless otherwise stated in a figure caption, each group of comparisons is performed on the same platform using identical affinity, compiler, and measurement settings.

\textbf{Workloads.} PCS, the Personal Communication System~\cite{luong2019applications}, is a communication-intensive benchmark in which agents represent communication nodes and generate five task types on a network topology. Aircraft is a compute-intensive UAV navigation and collision-avoidance benchmark using paths derived from real flight data~\cite{orr2023multiagent}. By varying entity scale, task complexity, and interaction frequency, the two parameterized workloads test SEER under different event granularities and contention patterns. 

\textbf{Comparison systems.} SEER w/o STPA enables the dual-layer event queue and dynamic parallel scheduling while using a manually specified thread-pool size. SEER denotes the complete system, which additionally enables STPA to select the thread-pool size from preflight traces and static workload features. We compare them with ROSS-T, which statically binds entities to threads; NBSE, which uses one global queue; and ROOT-Sim, which uses optimistic execution and dynamic rebinding~\cite{fujimoto1990parallel,jefferson1985virtual,michael1996simple}. All comparisons within the same figure use the same workload configuration and hardware platform.

\textbf{Metric definitions.}
Environment-execution throughput is reported as the average number of completed environment steps per second. Unless otherwise specified, speedup is computed by normalizing the throughput of a parallel configuration to that of the corresponding single-thread configuration under the same workload, platform, event handlers, and scenario parameters:
\begin{equation}
\mathrm{Speedup}(m)=
\frac{\mathrm{Throughput}(m\ \mathrm{threads})}
{\mathrm{Throughput}(1\ \mathrm{thread})}.
\end{equation}
Therefore, absolute throughput values such as 16.92 and 41.22 steps/s should not be interpreted as speedup ratios. The motivation study in Fig.~\ref{fig:3} uses a different workload configuration and directly compares end-to-end inference throughput between ROSS-T and MAgent2; its 1.03--1.22$\times$ ratios are therefore not directly comparable with the absolute environment-execution throughput reported in the main experimental evaluation.

\textbf{Measurement protocol.}
Within each comparison, all systems use the same event handlers, event-generation logic, policy network, action traces, compiler settings, CPU-affinity policy, and measurement windows. We report the aggregate throughput and latency measurements obtained from the evaluated configurations. Thread-prediction accuracy is evaluated separately using the training and test errors reported in Table~\ref{tab:thread_prediction}.

\subsection{Overall Performance}

\begin{figure}[!t]
    \centering
    \includegraphics[width=1\linewidth]{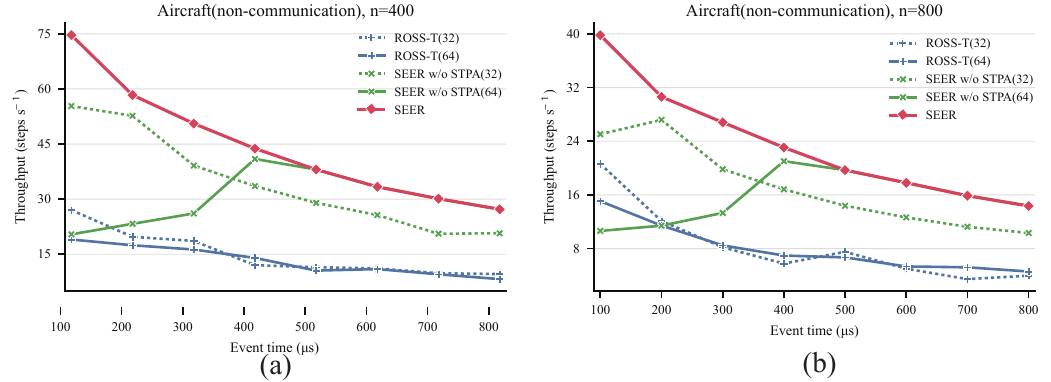}
    \caption{Absolute environment-execution throughput of ROSS-T, SEER w/o STPA, and SEER under different event granularities in the Aircraft workload on Platform A. Throughput is measured in environment steps per second; the reported values are not speedup ratios.}
    \label{fig:aap1}
\end{figure}

On Aircraft, average environment-execution throughput increases from 16.92 steps/s with ROSS-T to 41.22 steps/s with SEER w/o STPA and 52.93 steps/s with SEER (Fig.~\ref{fig:aap1}). Dynamic event sharing therefore provides a 143.63\% gain, while STPA adds 28.41\%.

\begin{figure}[!t]
    \centering
    \includegraphics[width=1\linewidth]{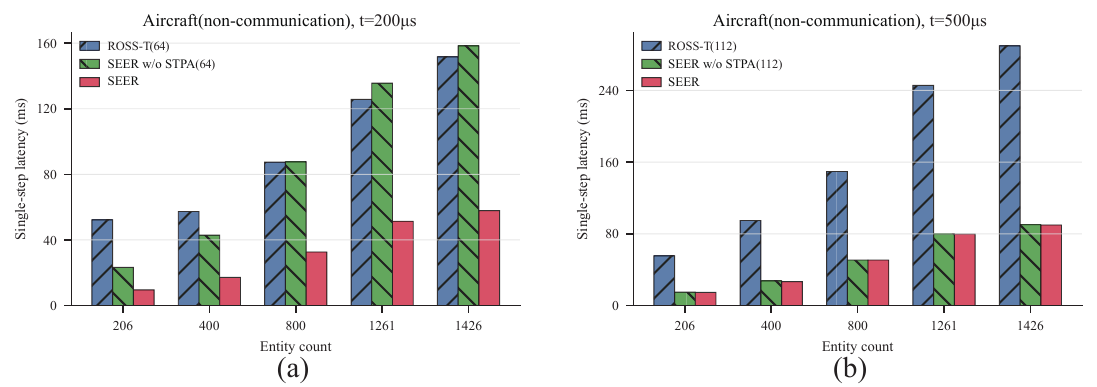}
    \caption{Environment-execution latency of ROSS-T, SEER w/o STPA, and SEER under different entity scales in the Aircraft workload on Platform A.}
    \label{fig:aap2}
\end{figure}

Across 10 Aircraft configurations, SEER reduces latency by 68.32\% over ROSS-T and 31.17\% over SEER w/o STPA (Fig.~\ref{fig:aap2}), consistent with the roles of dynamic event sharing and thread selection.

\begin{figure*}[!t]
    \centering
    \includegraphics[width=1\linewidth]{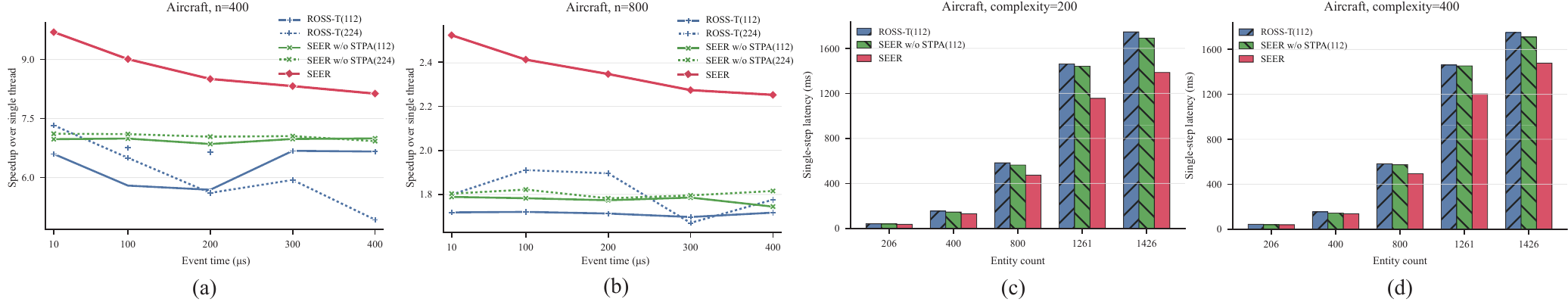}
   \caption{MARL environment-execution performance for Aircraft: (a--b) single-thread-normalized speedup versus average event time, and (c--d) single-step environment-execution latency versus entity scale. Speedup is computed relative to the corresponding single-thread throughput under the same workload and platform configuration.}
    \label{fig:aap3}
\end{figure*}

Figure~\ref{fig:aap3} shows that physical- or logical-core-sized pools are not consistently optimal. With 800 entities and $400\,\mu\mathrm{s}$ events, SEER achieves a 2.81$\times$ single-thread speedup using four workers. Across the measured configurations, STPA improves speedup by 14.52\% and reduces latency by 16.17\% over SEER w/o STPA.

\begin{figure}[!t]
    \centering
    \includegraphics[width=1\linewidth]{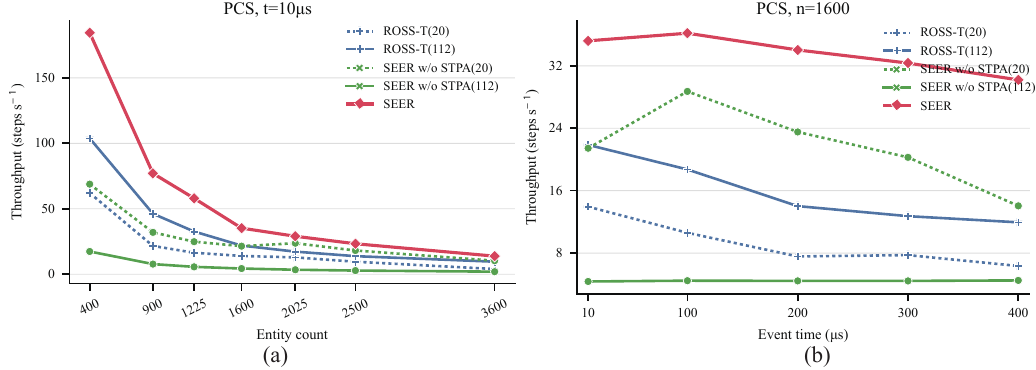}
    \caption{End-to-end MARL inference throughput of PCS on Platform B under varying event granularities and entity scales. Throughput is measured in inference steps per second.}
    \label{fig:aap4}
\end{figure}

End-to-end PCS results confirm that thread selection affects the full MARL loop (Fig.~\ref{fig:aap4}). At $n=3600$ and $t=10\,\mu\mathrm{s}$, SEER w/o STPA with 20 threads reaches 5.3$\times$ its 112-thread throughput; the corresponding ROSS-T ratio is 2.4$\times$.

\subsection{Load Balance and Thread Prediction}

\begin{figure}[!t]
    \centering
    \includegraphics[width=1\linewidth]{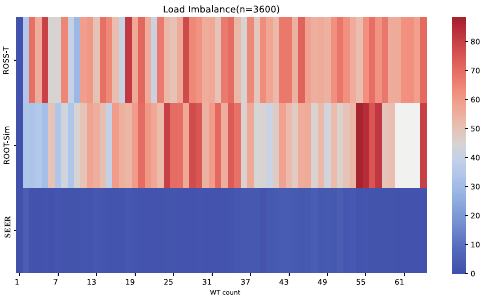}
   \caption{Per-thread CPU-time distribution of ROSS-T, ROOT-Sim, and SEER in the evaluated PCS configuration. A more uniform distribution across worker threads indicates better load balance.}
    \label{fig:13}
\end{figure}

\textbf{Sources of load balancing.}
In PCS with 3600 entities and $10\,\mu\mathrm{s}$ events, CPU-utilization variance drops from 23.79 with ROSS-T to 2.59 with SEER w/o STPA. SEER also improves single-thread-normalized speedup by 44.4\% over ROSS-T and 33.8\% over SEER w/o STPA. The per-thread distribution in Fig.~\ref{fig:13} supports the intended effect of local queues and global ready-work sharing.


\textbf{Thread selection.}
The useful thread count depends on event granularity, interaction intensity, and platform resources. In PCS with 3600 entities and $10\,\mu\mathrm{s}$ events, STPA's 20-thread choice yields 5.3$\times$ the NBSE throughput of the physical-core configuration. Thread scans show a measured rise-then-fall trend, motivating preflight-guided selection.

Table~\ref{tab:thread_prediction} compares STPA with TimesURL, XGBoost, LSTM, and TR+GP. STPA has the lowest test MAE, 0.3394, although the gap from its training MAE of 0.0108 indicates reduced generalization to unseen configurations.

\begin{table}[htbp]
\centering
\caption{Thread-count prediction errors of the evaluated thread-pool selection methods.}
\label{tab:thread_prediction}

\small
\begin{tabular}{lrrrr}
\hline
\textbf{Model} & \multicolumn{2}{c}{\textbf{Training Set}} & \multicolumn{2}{c}{\textbf{Test Set}} \\
& \textbf{MAE} & \textbf{MSE} & \textbf{MAE} & \textbf{MSE} \\
\hline
LSTM     & 8.9976  & 155.5296  & 9.3542   & 161.8647  \\
XGBoost  & 0.5525  & 0.6449    & 1.2832   & 5.1508    \\
TimesURL  & 1.8344  & 13.7815   & 2.2386   & 16.6097   \\
TR+GP    & 2.6535  & 18.3720   & 3.6187   & 31.2561   \\
STPA    & \textbf{0.0108} & \textbf{0.0109} &
            \textbf{0.3394} & \textbf{4.5944} \\
\hline
\end{tabular}
\end{table}

\subsection{Ablation Analysis}

ROSS-T uses static binding, SEER w/o STPA adds the dual-layer queue and dynamic scheduling, and SEER additionally enables STPA. Their average Aircraft throughputs are 16.92, 41.22, and 52.93 steps/s, respectively, showing that dynamic event sharing provides the larger gain and STPA adds 28.41\%. On PCS, SEER reduces latency by 46.9\%, 19.5\%, and 64.9\% over ROSS-T, NBSE, and ROOT-Sim. It uses 1.71~GB on average, about $1/47$ of ROOT-Sim and 0.44~GB more than ROSS-T.

\section{Related Work}

\textbf{Parallel MARL environments.}
Vectorized runtimes such as EnvPool parallelize execution across independent environment instances~\cite{weng2022envpool}. SEER instead targets event-level parallelism among interacting entities within one shared environment step; the two approaches are complementary and rely on different execution abstractions.

This work relates to two lines of research: load balancing in event-driven execution and resource configuration for parallel runtimes.

\textbf{Load balancing.}
Load balancing is a central problem in event-driven execution. Existing runtimes often bind logical processes or environment entities to processing units, and then rely on object migration~\cite{trabes2023parallel,cardenas2025lock,8} or dynamic regrouping~\cite{r21} to rebalance work. These designs are effective when migration is cheap, but they can be costly when entity state is large or frequently updated. Other studies improve load sharing through spatial or temporal locality~\cite{r4}, adaptive synchronization protocols~\cite{eker2021load,piccione2023hybrid}, fine-grained task scheduling~\cite{faxen2010efficient,wan2025parallel}, or lock-free data structures~\cite{marotta2024conflict,contini2023enabling,yang2023optimizing}. SEER differs by treating environment events, rather than entities, as the scheduling unit. SEER uses a distributed local-queue plus global-ready-list structure to preserve per-entity order and share ready work across threads.

\textbf{Resource configuration.}
Thread-pool size strongly affects parallel environment-execution performance, but existing event-driven runtimes provide limited support for flexible configuration across MARL workloads~\cite{r5,r6}. Setting the thread count equal to the number of physical cores~\cite{lorenzon2019parallel,shoshany2024c++} is simple, but it can be suboptimal when event granularity or interaction intensity changes. Performance modeling techniques~\cite{hafezan2021accurate,r13,r41} analyze the relation between thread count and execution time, but simplified models may miss queue contention and cache effects in event-driven workloads. Online tuning methods~\cite{r16,bahadur2025poolrunner,freire2021performance} can adjust thread counts, but iterative tuning is better suited to long-running services than to short environment steps. Profiling-based studies~\cite{chheda2024evaluating} and machine-learning approaches~\cite{gao2024autosched,r47} predict resource parameters from application features. SEER combines short preflight traces with static workload and hardware features, making thread-pool selection practical for the short formal executions considered in this paper.

\section{Conclusion}

This paper presented SEER, an event-driven runtime for reducing CPU-side environment-execution overhead in interaction-intensive MARL inference. SEER models an environment step as ordered events under an explicit event-locality contract, preserves correctness through entity-local causality, and exposes ready events to global worker threads through a CAS-based dual-layer queue. It further uses STPA to select a workload-aware thread scale from short preflight traces and static features. Across compute-intensive Aircraft and communication-intensive PCS workloads, SEER improved load balance, avoided excessive-thread contention, and accelerated environment execution while preserving the checked step outputs under the event-locality contract. These results suggest that single-step environment execution is an important systems target for scalable MARL inference.
\bibliography{bses_reference}

@String{Computing = "Computing" }

@String{Computer = "{IEEE} Computer" }

@String{Springer = "Springer-Verlag" }

@article{r4,
  title={Spatial/temporal locality-based load-sharing in speculative discrete event simulation on multi-core machines},
  author={Montesano, F and Marotta, R and Quaglia, F},
  journal={ACM Transactions on Modeling and Computer Simulation},
  volume={35},
  number={1},
  pages={1--31},
  year={2024}
}

@inproceedings{r5,
  title={The rome optimistic simulator: Core internals and programming model},
  author={Pellegrini, A and Vitali, R and Quaglia, F},
  booktitle={SIMUTools'11: Simulation Tools and Techniques},
  pages={96--98},
  year={2011}
}

@article{r6,
  title={ROSS: A high-performance, low-memory, modular Time Warp system},
  author={Carothers, CD and Bauer, D and Pearce, S},
  journal={Journal of Parallel and Distributed Computing},
  volume={62},
  number={11},
  pages={1648--1669},
  year={2002}
}

@inproceedings{r13, title={An Adaptive Approach For Parallel Discrete Event Simulation Thread Pool Prediction.}, author={Su, JiMing and Yao, Yiping and Zhu, Feng and Yuan, H and Fang, Y and Chen, K and Sui, S and Zhao, T and Zhang, H and Zhang, Y}, booktitle={ECMS}, pages={352--358}, year={2024} }

@article{r16,
  title={Towards optimal thread pool configuration for run-time systems of integration platforms},
  author={Freire, DL and Frantz, RZ and Roos-Frantz, F},
  journal={International Journal of Computer Applications in Technology},
  volume={62},
  number={2},
  pages={129--147},
  year={2020}
}

@article{r21,
  title={A discrete PSO-based static load balancing algorithm for distributed simulations in a cloud environment},
  author={Miao, Z and Yong, P and Mei, Y and others},
  journal={Future Generation Computer Systems},
  volume={115},
  pages={497--516},
  year={2021}
}

@article{r41,
  title={Predicting cross-core performance interference on multicore processors with regression analysis},
  author={Zhao, J and Cui, H and Xue, J and others},
  journal={IEEE Transactions on Parallel and Distributed Systems},
  volume={27},
  number={5},
  pages={1443--1456},
  year={2015}
}

@inproceedings{r47,
  title={Machine Learning Based Thread Pool Tuning via Program Analysis},
  author={Akash, L and Fernando, D and Jayasinghe, M and others},
  booktitle={2021 IEEE 23rd International Conference on High Performance Computing \& Communications; 7th International Conference on Data Science \& Systems; 19th International Conference on Smart City; 7th International Conference on Dependability in Sensor, Cloud \& Big Data Systems \& Application (HPCC/DSS/SmartCity/DependSys)},
  pages={648--653},
  year={2021}
}

@inproceedings{r48,
  title={TimesURL: Self-supervised contrastive learning for universal time series representation learning},
  author={Liu, J and Chen, S},
  booktitle={Proceedings of the AAAI Conference on Artificial Intelligence},
  volume={38},
  number={12},
  pages={13918--13926},
  year={2024}
}

@inproceedings{r49,
  title={A scalable tree boosting system},
  author={Chen, T and Guestrin, C},
  booktitle={Proceedings of the 22nd ACM SIGKDD International Conference on Knowledge Discovery and Data Mining},
  pages={785--794},
  year={2016}
}

@article{trabes2023parallel, title={A parallel algorithm to accelerate DEVS simulations in shared memory architectures}, author={Trabes, Guillermo German and Wainer, Gabriel A and Gil-Costa, Veronica}, journal={IEEE Transactions on Parallel and Distributed Systems}, volume={34}, number={5}, pages={1609--1620}, year={2023}, publisher={IEEE} }

@article{cardenas2025lock, title={Lock-Free Simulation Algorithm to Enhance the Performance of Sequential and Parallel DEVS Simulators in Shared-Memory Architectures}, author={C{\'a}rdenas, Rom{\'a}n and Arroba, Patricia and Risco-Mart{\'\i}n, Jos{\'e} L}, journal={Journal of Parallel and Distributed Computing}, pages={105105}, year={2025}, publisher={Elsevier} }

@inproceedings{8,
  title={Effective access to the committed global state in speculative parallel discrete event simulation on multi-core machines},
  author={Marotta, Romolo and Montesano, Federica and Quaglia, Francesco},
  booktitle={Proceedings of the 2023 ACM SIGSIM Conference on Principles of Advanced Discrete Simulation},
  pages={107--117},
  year={2023}
}

@inproceedings{eker2021load, title={Load-aware dynamic time synchronization in parallel discrete event simulation}, author={Eker, Ali and Arafa, Yehia and Badawy, Abdel-Hameed A and Santhi, Nandakishore and Eidenbenz, Stephan and Ponomarev, Dmitry}, booktitle={Proceedings of the 2021 ACM SIGSIM Conference on Principles of Advanced Discrete Simulation}, pages={95--105}, year={2021} }

@inproceedings{piccione2023hybrid, title={Hybrid speculative synchronisation for parallel discrete event simulation}, author={Piccione, Andrea and Andelfinger, Philipp and Pellegrini, Alessandro}, booktitle={Proceedings of the 2023 ACM SIGSIM Conference on Principles of Advanced Discrete Simulation}, pages={84--95}, year={2023} }

@inproceedings{faxen2010efficient, title={Efficient work stealing for fine grained parallelism}, author={Faxen, Karl-Filip}, booktitle={2010 39th International Conference on Parallel Processing}, pages={313--322}, year={2010}, organization={IEEE} }

@inproceedings{wan2025parallel, title={Parallel contraction hierarchies can be efficient and scalable}, author={Wan, Zijin and Dong, Xiaojun and Wang, Letong and Zhu, Enzuo and Gu, Yan and Sun, Yihan}, booktitle={Proceedings of the 39th ACM International Conference on Supercomputing}, pages={670--688}, year={2025} }

@article{marotta2024conflict, title={A conflict-resilient lock-free linearizable calendar queue}, author={Marotta, Romolo and Ianni, Mauro and Pellegrini, Alessandro and Quaglia, Francesco}, journal={ACM transactions on parallel computing}, volume={11}, number={1}, pages={1--32}, year={2024}, publisher={ACM New York, NY} }

@inproceedings{contini2023enabling, title={Enabling reconfigurable HPC through MPI-based inter-FPGA communication}, author={Contini, Nicholas and Ramesh, Bharath and Kandadi Suresh, Kaushik and Tran, Tu and Michalowicz, Ben and Abduljabbar, Mustafa and Subramoni, Hari and Panda, Dhabaleswar}, booktitle={Proceedings of the 37th International Conference on Supercomputing}, pages={477--487}, year={2023} }

@inproceedings{yang2023optimizing, title={Optimizing multi-grid computation and parallelization on multi-cores}, author={Yang, Xiaojian and Li, Shengguo and Yuan, Fan and Dong, Dezun and Huang, Chun and Wang, Zheng}, booktitle={Proceedings of the 37th International Conference on Supercomputing}, pages={227--239}, year={2023} }

@book{lorenzon2019parallel, title={Parallel computing hits the power wall: principles, challenges, and a survey of solutions}, author={Lorenzon, Arthur Francisco and Beck Filho, Antonio Carlos Schneider}, year={2019}, publisher={Springer Nature} }

@article{shoshany2024c++, title={A {C++} 17 thread pool for high-performance scientific computing}, author={Shoshany, Barak}, journal={SoftwareX}, volume={26}, pages={101687}, year={2024}, publisher={Elsevier} }

@inproceedings{hafezan2021accurate, title={An Accurate Learning-Based Performance/Power Model for System-Level Design of a Multicore Multithreaded Network Processor}, author={Hafezan, Mohamad and Azari, Hossein and Dabaghan, Amir and Beigi, Leila}, booktitle={2021 4th International Conference on Information and Computer Technologies (ICICT)}, pages={149--157}, year={2021}, organization={IEEE} }

@inproceedings{popov2019efficient, title={Efficient thread/page/parallelism autotuning for numa systems}, author={Popov, Mihail and Jimborean, Alexandra and Black-Schaffer, David}, booktitle={Proceedings of the ACM International Conference on Supercomputing}, pages={342--353}, year={2019} }

@article{bahadur2025poolrunner, title={PoolRunner: An Extensible Performance Testing Simulation Tool for Thread-Pool Middleware}, author={Bahadur, Faisal and Ahmad, Zulfiqar and Algarni, Abdulmohsen}, journal={IEEE Access}, year={2025}, publisher={IEEE} }

@article{freire2021performance, title={Performance evaluation of thread pool configurations in the run-time systems of integration platforms}, author={Freire, Daniela L and Mazzonetto, Angela and Frantz, Rafael Z and Roos-Frantz, Fabricia and Sawicki, Sandro and Basto-Fernandes, Vitor}, journal={International Journal of Business Process Integration and Management}, volume={10}, number={3-4}, pages={318--329}, year={2021}, publisher={Inderscience Publishers (IEL)} }

@inproceedings{chheda2024evaluating, title={Evaluating Tuning Opportunities of the LLVM/OpenMP Runtime}, author={Chheda, Smeet and Verma, Gaurav and Tian, Shilei and Chapman, Barbara and Doerfert, Johannes}, booktitle={SC24-W: Workshops of the International Conference for High Performance Computing, Networking, Storage and Analysis}, pages={919--929}, year={2024}, organization={IEEE} }

@inproceedings{gao2024autosched, title={Autosched: An adaptive self-configured framework for scheduling deep learning training workloads}, author={Gao, Wei and Zhang, Xu and Huang, Shan and Guo, Shangwei and Sun, Peng and Wen, Yonggang and Zhang, Tianwei}, booktitle={Proceedings of the 38th ACM International Conference on Supercomputing}, pages={473--484}, year={2024} }

@article{vinyals2019grandmaster,
  title = {Grandmaster level in {StarCraft II} using multi-agent reinforcement learning},
  author = {Vinyals, Oriol and Babuschkin, Igor and Czarnecki, Wojciech M. and Mathieu, Micha{\"e}l and Dudzik, Andrew and Chung, Junyoung and Choi, David H. and Powell, Richard and Ewalds, Timo and Georgiev, Petko and Oh, Junhyuk and Horgan, Dan and Kroiss, Manuel and Danihelka, Ivo and Huang, Aja and Sifre, Laurent and Cai, Trevor and Agapiou, John P. and Jaderberg, Max and Vezhnevets, Alexander S. and Leblond, R{\'e}mi and Pohlen, Tobias and Dalibard, Valentin and Budden, David and Sulsky, Yury and Molloy, James and Paine, Tom L. and Gulcehre, Caglar and Wang, Ziyu and Pfaff, Tobias and Wu, Yuhuai and Ring, Roman and Yogatama, Dani and W{\"u}nsch, Dario and McKinney, Katrina and Smith, Oliver and Schaul, Tom and Lillicrap, Timothy and Kavukcuoglu, Koray and Hassabis, Demis and Apps, Chris and Silver, David},
  journal = {Nature},
  volume = {575},
  number = {7782},
  pages = {350--354},
  year = {2019},
  doi = {10.1038/s41586-019-1724-z},
  url = {https://doi.org/10.1038/s41586-019-1724-z}
}

@article{jaderberg2019human,
  title = {Human-level performance in 3D multiplayer games with population-based reinforcement learning},
  author = {Jaderberg, Max and Czarnecki, Wojciech M. and Dunning, Iain and Marris, Luke and Lever, Guy and Casta{\~n}eda, Antonio Garcia and Beattie, Charles and Rabinowitz, Neil C. and Morcos, Ari S. and Ruderman, Avraham and Sonnerat, Nicolas and Green, Tim and Deason, Louise and Leibo, Joel Z. and Silver, David and Hassabis, Demis and Kavukcuoglu, Koray and Graepel, Thore},
  journal = {Science},
  volume = {364},
  number = {6443},
  pages = {859--865},
  year = {2019},
  doi = {10.1126/science.aau6249},
  url = {https://doi.org/10.1126/science.aau6249}
}

@inproceedings{terry2021pettingzoo,
  title = {{PettingZoo}: Gym for multi-agent reinforcement learning},
  author = {Terry, Justin K. and Black, Benjamin and Grammel, Nathaniel and Jayakumar, Mario and Hari, Ananth and Sullivan, Ryan and Santos, Luis and Perez, Rodrigo and Horsch, Caroline and Dieffendahl, Clemens and Williams, Niall L. and Lokesh, Yashas and Ravi, Praveen},
  booktitle = {Advances in Neural Information Processing Systems},
  volume = {34},
  year = {2021},
  doi = {10.48550/arXiv.2009.14471},
  url = {https://arxiv.org/abs/2009.14471}
}

@article{zheng2018magent,
  title = {{MAgent}: A many-agent reinforcement learning platform for artificial collective intelligence},
  author = {Zheng, Lianmin and Yang, Jiacheng and Cai, Han and Zhou, Ming and Zhang, Weinan and Wang, Jun and Yu, Yong},
  journal = {Proceedings of the AAAI Conference on Artificial Intelligence},
  volume = {32},
  number = {1},
  year = {2018},
  doi = {10.1609/aaai.v32i1.11371},
  url = {https://doi.org/10.1609/aaai.v32i1.11371}
}

@inproceedings{yu2022surprising,
  title = {The surprising effectiveness of {PPO} in cooperative multi-agent games},
  author = {Yu, Chao and Velu, Akash and Vinitsky, Eugene and Gao, Jiaxuan and Wang, Yu and Bayen, Alexandre and Wu, Yi},
  booktitle = {Advances in Neural Information Processing Systems},
  volume = {35},
  year = {2022},
  doi = {10.48550/arXiv.2103.01955},
  url = {https://arxiv.org/abs/2103.01955}
}

@inproceedings{weng2022envpool,
  title = {{EnvPool}: A highly parallel reinforcement learning environment execution engine},
  author = {Weng, Jiayi and Lin, Min and Huang, Shengyi and Liu, Bo and Makoviichuk, Denys and Makoviychuk, Viktor and Liu, Zichen and Song, Yufan and Luo, Ting and Jiang, Yukun and Xu, Zhongwen and Yan, Shuicheng},
  booktitle = {Advances in Neural Information Processing Systems},
  volume = {35},
  year = {2022},
  doi = {10.48550/arXiv.2206.10558},
  url = {https://arxiv.org/abs/2206.10558}
}

@article{fujimoto1990parallel,
  title = {Parallel discrete event simulation},
  author = {Fujimoto, Richard M.},
  journal = {Communications of the ACM},
  volume = {33},
  number = {10},
  pages = {30--53},
  year = {1990},
  doi = {10.1145/84537.84545},
  url = {https://doi.org/10.1145/84537.84545}
}

@article{jefferson1985virtual,
  title = {Virtual time},
  author = {Jefferson, David R.},
  journal = {ACM Transactions on Programming Languages and Systems},
  volume = {7},
  number = {3},
  pages = {404--425},
  year = {1985},
  doi = {10.1145/3916.3988},
  url = {https://doi.org/10.1145/3916.3988}
}

@inproceedings{michael1996simple,
  title = {Simple, fast, and practical non-blocking and blocking concurrent queue algorithms},
  author = {Michael, Maged M. and Scott, Michael L.},
  booktitle = {Proceedings of the Fifteenth Annual ACM Symposium on Principles of Distributed Computing},
  pages = {267--275},
  year = {1996},
  doi = {10.1145/248052.248106},
  url = {https://doi.org/10.1145/248052.248106}
}

@article{blumofe1999scheduling,
  title = {Scheduling multithreaded computations by work stealing},
  author = {Blumofe, Robert D. and Leiserson, Charles E.},
  journal = {Journal of the ACM},
  volume = {46},
  number = {5},
  pages = {720--748},
  year = {1999},
  doi = {10.1145/324133.324234},
  url = {https://doi.org/10.1145/324133.324234}
}

@article{hill2008amdahl,
  title = {Amdahl's Law in the multicore era},
  author = {Hill, Mark D. and Marty, Michael R.},
  journal = {Computer},
  volume = {41},
  number = {7},
  pages = {33--38},
  year = {2008},
  doi = {10.1109/MC.2008.209},
  url = {https://doi.org/10.1109/MC.2008.209}
}

@inproceedings{pusukuri2011thread,
  title = {Thread reinforcer: Dynamically determining number of threads via {OS} level monitoring},
  author = {Pusukuri, Kishore Kumar and Gupta, Rajiv and Bhuyan, Laxmi N.},
  booktitle = {2011 IEEE International Symposium on Workload Characterization},
  year = {2011},
  doi = {10.1109/IISWC.2011.6114208},
  url = {https://doi.org/10.1109/IISWC.2011.6114208}
}

@article{alcaraz2022predicting,
  title = {Predicting number of threads using balanced datasets for {OpenMP} regions},
  author = {Alcaraz, Jordi and TehraniJamsaz, Ali and Dutta, Akash and Sikora, Anna and Jannesari, Ali and Sorribes, Joan and Cesar, Eduardo},
  journal = {Computing},
  year = {2022},
  doi = {10.1007/s00607-022-01081-6},
  url = {https://doi.org/10.1007/s00607-022-01081-6}
}

@article{luong2019applications,
  title = {Applications of deep reinforcement learning in communications and networking: A survey},
  author = {Luong, Nguyen Cong and Hoang, Dinh Thai and Gong, Shimin and Niyato, Dusit and Wang, Ping and Liang, Ying-Chang and Kim, Dong In},
  journal = {IEEE Communications Surveys \& Tutorials},
  volume = {21},
  number = {4},
  pages = {3133--3174},
  year = {2019},
  doi = {10.1109/COMST.2019.2916583},
  url = {https://doi.org/10.1109/COMST.2019.2916583}
}

@article{orr2023multiagent,
  title = {Multi-agent deep reinforcement learning for multi-robot applications: A survey},
  author = {Orr, James and Dutta, Ayan},
  journal = {Sensors},
  volume = {23},
  number = {7},
  pages = {3625},
  year = {2023},
  doi = {10.3390/s23073625},
  url = {https://doi.org/10.3390/s23073625}
}

@inproceedings{jagtap2012optimization,
  title={Optimization of parallel discrete event simulator for multi-core systems},
  author={Jagtap, Deepak and Abu-Ghazaleh, Nael and Ponomarev, Dmitry},
  booktitle={2012 IEEE 26th International Parallel and Distributed Processing Symposium},
  pages={520--531},
  year={2012},
  organization={IEEE}
}
\end{document}